\documentclass{Interspeech}

\interspeechcameraready

\usepackage{cite}
\usepackage{amsmath,amssymb,amsfonts}
\usepackage{algorithmic}
\usepackage{graphicx}
\usepackage{textcomp}

\usepackage{booktabs}
\usepackage{hyperref}
\usepackage{IEEEtrantools}

\usepackage{multirow,float,enumitem}
\usepackage{bibspacing}
\usepackage{bbding}
\usepackage{pifont}

\newcommand{\hnxu}[1]{{\color{black}{#1}}}
\newcommand{\hnxured}[1]{{\color{black}{#1}}}
\newcommand{\tim}[1]{\textcolor{black}{#1}}
\newcommand{\zq}[1]{\textcolor{black}{#1}}

\newcommand{\timnew}[1]{\textcolor{black}{#1}}

\newcommand{\hnmod}[1]{\textcolor{black}{#1}}
\usepackage[table]{xcolor}

\title{Effective and Efficient One-pass Compression of Speech Foundation Models Using Sparsity-aware Self-pinching Gates}
\author[affiliation={1}]{Haoning}{Xu}
\author[affiliation={1}]{Zhaoqing}{Li}
\author[affiliation={1}]{Youjun}{Chen}
\author[affiliation={1}]{Huimeng}{Wang}
\author[affiliation={1}]{Guinan}{Li}
\author[affiliation={2}]{Mengzhe}{Geng}
\author[affiliation={1}]{Chengxi}{Deng}
\author[affiliation={1}]{Xunying}{Liu}

\affiliation{}{The Chinese University of Hong Kong}{Hong Kong SAR, China}
\affiliation{}{National Research Council Canada}{Canada}
\email{hnxu@se.cuhk.edu.hk, xyliu@se.cuhk.edu.hk}

\usepackage{comment}

\begin{document}
\bstctlcite{IEEEexample:BSTcontrol}
\maketitle

\begin{abstract}
This paper presents a novel approach for speech foundation models compression that tightly integrates model pruning and parameter update into a single stage. Highly compact layer-level tied self-pinching gates each containing only a single learnable threshold are jointly trained with uncompressed models and used in fine-grained neuron level pruning. Experiments conducted on the LibriSpeech-100hr corpus suggest that our approach reduces the number of parameters of wav2vec2.0-\textit{base} and HuBERT-\textit{large} models by 65\% and 60\% respectively, while incurring no statistically significant word error rate (WER) increase on the test-clean dataset. Compared to previously published methods on the same task, our approach not only achieves the lowest WER of 7.05\% on the test-clean dataset under a comparable model compression ratio of 4.26x, but also operates with at least 25\% less model compression time. 

    
\end{abstract}

\section{Introduction}

In recent years, advancements in self-supervised learning (SSL) for speech technologies, particularly through foundation models like wav2vec2.0~\cite{baevski2020wav2vec}, HuBERT~\cite{hsu2021HuBERT} and WavLM~\cite{chen2022wavlm}, have greatly improved their utility in applications like automatic speech recognition (ASR). Despite these advancements, the widespread adoption of these models in resource-constrained and on-device environments is limited due to their substantial memory and computational demands.

To address this challenge, extensive research has explored diverse neural network compression methods for ASR tasks, including but not limited to: \textbf{1) low-bit quantization} approaches that reduce memory footprint by replacing floating-point weights with \tim{low-precision} values~\cite{uq-2bcfm,uq-4bcfm,ibert,xu2025effective,mq-person}; and \textbf{2) architecture compression} methods that focus on reducing structural redundancy in models, such as low-rank matrix factorization~\cite{lr3,li2023lossless,wang2024fact}, knowledge distillation~\cite{rathod2022multi,park2023conformer,distw2v,distilhubert,wang2022lighthubert,deepvswide,fithubert} and model pruning~\cite{pru3,pru5,gu2024sparsewav,wang2023task,jiang2023accurate,lodagala2023pada,hj}. Furthermore, larger model compression ratios can be achieved by combining low-bit quantization and architecture compression~\cite{li2023lossless,onepass-zq,mp-w2v,usmlite}.

However, previous studies on model pruning of \timnew{SSL-based} ASR systems 
\timnew{face} the following limitations: 
%
\textbf{1) Significant performance degradation} is often observed 
\tim{as an increase in WER} after performing compression. 
\tim{Coarse-grained ``structured'' pruning methods, which operate at the level of larger structures (e.g., channels) rather than individual parameters, may inadvertently remove critical parameters, resulting in performance degradation~\cite{wang2023task,peng2023dphubert,hj,zampierin2024skillsimilarityawareknowledgedistillation}.}
It is also important to note that most studies failed to clearly define the criterion, such as statistical significance tests\cite{gillick1989some}, to differentiate ``acceptable" and ``unacceptable" performance loss due to compression. 
\textbf{2) Inconsistency between model pruning and parameter update} creates two separate and disjointed stages during compression, often leading to a large performance degradation. 
Approaches adopted in~\cite{lai2021parp, yang2023learning,lodagala2023pada,gu2024sparsewav} employ sequential paradigms that optimize pruning masks and parameters separately, which decouple pruning from parameter optimization \tim{in} the ASR task. For example, the parameter pruning in~\cite{gu2024sparsewav} is driven by an independent layer-wise evaluation before post-pruning fine-tuning. 
\textbf{3) Substantial training time} inevitably arises from post-pruning refinement stages, including knowledge distillation~\cite{peng2023dphubert}, fine-tuning~\cite{gu2024sparsewav}, and iterative pruning~\cite{lai2021parp,yang2023learning,lodagala2023pada,siyuan,mp-w2v}, which not only prolong development cycles but also complicate experimental workflows 
\tim{due to} multi-stage requirements.
\hnxured{\textbf{4) Excessive pruning-required parameter overhead}\tim{, such as}: 
\textbf{i}) the extra parameters in KD-based pruning~\cite{wang2023task,peng2023dphubert},
\tim{\textbf{ii)}} \tim{the} use of candidate architecture-specific weights in~\cite{li2023lossless, hj, wang2023task,peng2023dphubert} and 
\textbf{iii)} \tim{the parametrization of entire masks as trainable matrices~\cite{fu2022losses}}.}


To this end, this paper 
\timnew{proposes} a novel compression approach for SSL speech foundation models that tightly integrates model pruning and parameter update into a single stage, \hnmod{referred to as the \textit{one-pass} stage.}
In each layer, the sparsity-aware self-pinching gate generates a differentiable pruning probability for each parameter by comparing its magnitude with the learnable threshold. For wav2vec2.0-\textit{base} and HuBERT-\textit{large}, 
only $6\times12=72$ and $6\times24=144$ additional parameters are introduced 
as \timnew{thresholds}, respectively\footnote{Compared to the uncompressed wav2vec2.0-\textit{base} and HuBERT-\textit{large} models,
the additional pruning-required parameters account for only 8e-7 and 5e-7 of the total model size, respectively.}. 

Experiments conducted on Librispeech dataset suggest that the pruned wav2vec2.0-\textit{base} and HuBERT-\textit{large} using our method \textbf{1)} significantly outperform 
\timnew{both \textbf{i})} uniform magnitude-based pruning \hnxured{where pruning and parameter update are separated, and \timnew{\textbf{ii)}} neural architecture search (NAS)~\cite{liu2018darts} based channel-wise pruning
\timnew{, which introduces additional parameters for coarse-grained pruning}}, \timnew{particularly} when the overall sparsity reaches 50\% or greater, and \textbf{2)} further achieve lossless\footnote{``lossless'' in this paper refers to no statistically significant WER increase against the uncompressed baseline.} compression with sparsity of 65\% and 60\% on the test-clean subset,
respectively. 
It should be noted that the time to compress wav2vec2.0-\textit{base} and HuBERT-\textit{large} only takes 11 and 13 GPU hours, respectively. Compared with prior pruning methods for SSL-based ASR systems, the proposed method obtains the lowest WER of 7.05\% on the test-clean subset, while also reducing fine-tuning time 
with the model size limited to 23M parameters
under a model compression ratio of 4.26x.

The main contributions of this paper include:

    \textbf{1)} 
    Our method reduces the fragility to pruning \timnew{by using} the proposed layer-level sparsity-aware self-pinching gate.
    Compared to the coarse-grained pruning approaches~\cite{wang2023task,peng2023dphubert,hj,zampierin2024skillsimilarityawareknowledgedistillation}, 
    our fine-grained method 
    assigns a specific pruning probability to each parameter, 
    leading to better ASR performance.

    \textbf{2)} Our method \timnew{ensures} consistency between model pruning and parameter update by 
    \timnew{integrating them into}
    a single stage for \timnew{SSL-based} ASR systems. \timnew{In contrast,} previous fine-grained 
    approach~\cite{gu2024sparsewav} \tim{performed pruning and fine-tuning separately}. \hnxured{This one-pass compression stage also enables different layers to be pruned in \timnew{varying} sparsity based on their sensitivities,
    \timnew{thereby achieving optimal mixed-sparsity assignments.}} 
    
    \textbf{3)} Our method \timnew{demonstrates} 
    efficiency 
    \timnew{in terms of} compression time, 
    \timnew{as it eliminates the need for} additional operations such as post-pruning fine-tuning~\cite{gu2024sparsewav} or iterative pruning~\cite{lai2021parp} after the one-pass compression stage.

    \textbf{4)} Our method guarantees the compactness of pruning-required parameters by introducing 
    a single threshold as an additional component for each layer. In contrast, the number of additional pruning-required parameters in previous methods is based on \textbf{i)} the design of teacher-student model~\cite{peng2023dphubert,wang2023task}, \textbf{ii)} the number of candidates~\cite{li2023lossless, hj} or \textbf{iii)} the layer size~\cite{fu2022losses}.
\vspace{-0.3cm}
\section{wav2vec2.0 and HuBERT Models}
Speech SSL models such as wav2vec2.0~\cite{baevski2020wav2vec}, HuBERT~\cite{hsu2021HuBERT}, and WavLM~\cite{chen2022wavlm} share similar Transformer backbones. 
For example, HuBERT consists of a CNN encoder, a Transformer encoder, a projection layer and
\timnew{a code embedding layer}. Transformer encoder accounts for over 90\% of the total number of parameters, where each encoder block contains an MHSA module and an FFN module. In this work, we fine-tuned wav2vec2.0-\textit{base} and HuBERT-\textit{large} with a pure CTC decoder and pruned parameters in the 6 linear layers of each Transformer encoder block.

\vspace{-0.3cm}
\section{Previous works} \label{Pre}

\subsection{Uniform Magnitude-based Pruning} \label{MP}
%
The Uniform Magnitude-based Pruning (UMP) is 
\timnew{inspired by} the 
\timnew{principle} that 
parameters \timnew{with smaller} \textbf{magnitudes (absolute values)} \timnew{have less influence on} the output.
As shown in Fig.~\ref{frame} (a), 
in UMP, the parameters of a specific layer are sorted by magnitude, and the same proportion of parameters with relatively small magnitude 
\timnew{is} pruned 
\timnew{across all layers}, 
leading to consistent sparsity throughout the network. 

However, UMP enforces isotropic sparsity across layers, ignoring layer-specific sensitivity profiles. 
\timnew{While some} layers tolerate aggressive pruning with 
\timnew{little} accuracy loss, others degrade significantly even with conservative sparsity.


\vspace{-0.2cm}
\subsection{NAS-based Channel-wise Pruning} \label{NAS}
 A \textbf{channel} is defined as a row (${index,:}$) or column (${:,index}$) of the weight matrix of a linear layer. For example, in the $p$-th Transformer block, \textbf{1)} the $m$-th channels in the \hnmod{weight matrices of Multi-Head Self-Attention} module,
 denoted as ${\{\mathbf{Q}\text{uery}_{m,:}^p, \mathbf{K}\text{ey}_{m,:}^p, \mathbf{V}\text{alue}_{m,:}^p, \mathbf{O}\text{ut}_{:,m}^p\}}$, are simultaneously pruned and \textbf{2)} the $n$-th channels in the \hnmod{weight matrices of \textbf{F}eed-\textbf{F}orward \textbf{N}etwork module}, denoted as ${\{\mathbf{FFN.1}_{n,:}^p}, {\mathbf{FFN.2}_{:,n}^p\}}$, are pruned together. The channels are sorted by the \textbf{sum of L2-magnitude}, which is calculated as $(\lVert{\mathbf{Q}^p_{m,:}}\rVert_{2}+ \lVert{\mathbf{K}^p_{m,:}}\rVert_{2}+\lVert{\mathbf{V}^p_{m,:}}\rVert_{2}+\lVert{\mathbf{O}^p_{:,m}}\rVert_{2})$ and $(\lVert{\mathbf{FFN.1}^p_{n,:}}\rVert_{2}+ \lVert{\mathbf{FFN.2}^p_{:,n}}\rVert_{2})$ for the $m$-th and $n$-th channels, respectively. Here $\lVert{\cdot}\rVert_{2}$ indicates the L2-norm. 

In the Gumbel-Softmax differentiable neural architecture search (DARTS)~\cite{maddison2022concrete,liu2018darts}, we train a supernet that simultaneously contains 7 \textbf{candidate masks} per layer, as shown in Fig.~\ref{frame} (b), where the channels with relatively small \textbf{sum of L2-magnitude} are pruned in proportions of ${\{0\%, 25\%, 50\%, 75\%, 87.5\%, 90\%, 92.5\%\}}$ of the total channels, respectively. 
Specifically, the masked weight matrix of the $l$-th layer $\hat{\mathbf{W}}^l$ can be computed as follows:
\vspace{-0.2cm}
\begin{equation}
\hat{\mathbf{W}}^l=\sum_{i=1}^7\lambda_i^l\cdot\overline{\mathbf{M}}_i^l\odot\mathbf{W}^{l},
\end{equation}
where $\mathbf{W}^{l}$ is the weight matrix, $\overline{\mathbf{M}}_i^l$ denotes the $i$-th candidate mask and $\lambda_i^l$ is the parameter that measures its importance.
Gumbel-softmax is obtained by Gumbel sampling the output of the Softmax function, which is given by:
\begin{equation}
\lambda_i^l=\frac{\exp((\log\alpha_i^l+G_i^l)/T)}{\sum_{j=1}^7\exp((\log\alpha_j^l+G_j^l)/T)},
\end{equation}
where $\alpha_i^l$ is an \textbf{architecture-dependent parameter} determining their contribution during search. $G_i^l = -\log(-\log(U_i^l))$ is the Gumbel variable, $T$ is a hyperparameter and $U_l^i$ is a uniform random variable. 
Let $\lVert{\cdot}\rVert_{0}$ indicate the number of non-zero parameters, the training loss is designed as:
\vspace{-0.2cm}
\begin{equation}
    \mathcal{L} =\mathcal{L}_{ctc} +\eta\sum_{l}{\sum_{i=1}^7\lambda_i^l\cdot\lVert{\overline{\mathbf{M}}}_i^l\rVert_{0}},\label{Pass 1 eq}
\end{equation}
where $\mathcal{L}_{ctc}$ is the CTC loss of the pruned system with 
\timnew{the overall sparsity dynamically searched in real time}
and $\eta$ is a constant coefficient that controls the overall sparsity. 

\timnew{While} NAS-based techniques ensure consistency between model pruning and parameter update, they introduce considerable computational and parameter overhead. Furthermore, their effectiveness is highly dependent on the design of candidates.

\vspace{-0.3cm}
\section{Sparsity-aware Self-pinching Gates}

The core concept of \textbf{Sparsity-aware Self-pinching Gates} is to leverage the weights already being learned to construct the mask using only one additional learnable threshold per layer. 
Our approach facilitates flexible allocation of layer-wise sparsity across different layers based on their sensitivities, ultimately enabling the model to reach a specific size limit.


\begin{figure*}[h]
    \centering
    \includegraphics[scale=0.09]{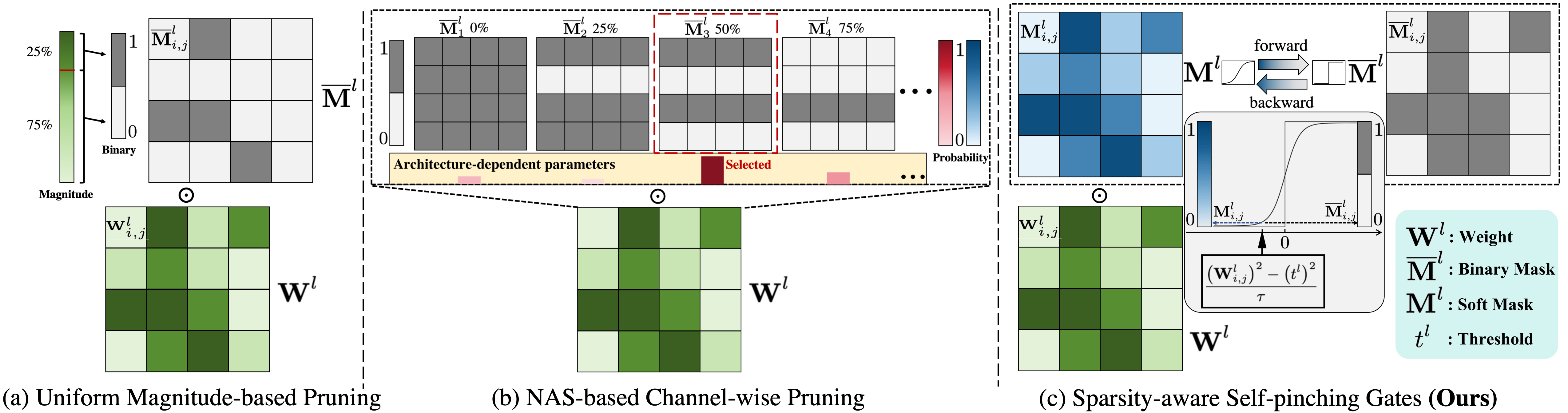}
    \vspace{-2mm}
    \caption{Comparison between Uniform Magnitude-based Pruning (UMP), NAS-based Channel-wise Pruning (NAS-CP) and Sparsity-aware Self-pinching Gates (ours). For the $l$-th layer, \textbf{(a)} UMP directly prunes the same proportion of parameters by magnitude \timnew{across} all layers; \textbf{(b)} NAS-CP introduces architecture-dependent parameters proportional to architecture candidates, which are pre-selected before NAS search; \textbf{(c)} Ours utilizes the weights that are already being learned to construct the mask with only one additional threshold.}
    \vspace{-5mm}
    \label{frame}
\end{figure*}

 For the $l$-th layer, $\mathbf{M}_{i, j}$ of the element-wise soft mask matrix $\mathbf{M}^{l}$ is expressed as:
\vspace{-0.2cm}
\begin{equation}
\mathbf{M}_{i, j}^{l}={\rm Sigmoid}\left ({\frac{\left(\mathbf{W}_{i, j}^{l}\right)^{2}-\left(t^{l}\right)^{2}}{\tau}} \right),
\end{equation}
where $t^l$ is a learnable threshold shared across all parameters in the weight matrix $\mathbf{W}^l$ of the $l$-th layer. Here, $i$ and $j$ denote the indices of parameters, and $\tau$ is a positive hyperparameter. Let $|\cdot|$ denote the number of parameters, the sparsity $\mathbf{s}^l$ of the $l$-th layer is defined as:
\vspace{-0.2cm}
\begin{equation}
\mathbf{s}^l=1-\frac{\lVert{\mathbf{M}}^l\rVert_{0}}{{|\mathbf{M}}^l|}.
\end{equation}



The Straight-Through Estimator (STE)~\cite{bengio2013estimatingSTE} is used during the one-pass fine-tuning stage: \textbf{1)} During forward propagation or inference, $\mathbf{M}^l$ is rounded to a binary mask $\overline{\mathbf{M}}^l$ and the masked weight matrix is computed as $\overline{\mathbf{W}}^{l}=\mathbf{W}^{l}\odot\overline{\mathbf{M}}^l$. Specifically, the parameters in $\mathbf{W}^{l}$ with magnitude exceeding the threshold $t^l$ will be retained, while those below the threshold will be pruned, which can be expressed as:
\vspace{-0.2cm}
\begin{equation}
\overline{\mathbf{M}}^l_{{i,j}}=\begin{cases}1&(\mathbf{W}^l_{i,j})^2\geq(t^l)^2\\0&(\mathbf{W}^l_{i,j})^2<(t^l)^2\end{cases},
\end{equation}
\textbf{2)} During backward propagation, the gradient is accumulated using the actual values in $\mathbf{M}^l$.
The training loss is given by:
\vspace{-0.2cm}
\begin{equation}
\mathcal{L} = \mathcal{L}_{ctc}+\eta\sum\limits_{l}\lVert{\overline{\mathbf{M}}}^l\rVert_{0},
\end{equation}
where $\eta$ is a constant coefficient to control the overall sparsity. $\eta$ is set to 0 if the desired sparsity is achieved; otherwise, it remains at its preset value.

Our method shown in Fig.~\ref{frame} (c) simultaneously optimizes the threshold $t^l$ and the weight matrix $\mathbf{W}^l$, thus mitigating the mismatch between model pruning and parameter update. 
\hnmod{
The above fine-grained neural level pruning is more powerful in mitigating performance loss due to model compression than coarse-grained pruning methods, which are applied at the channel level and easier to implement.
}
\begin{figure*}[h]
    \centering
    \includegraphics[scale=0.11]{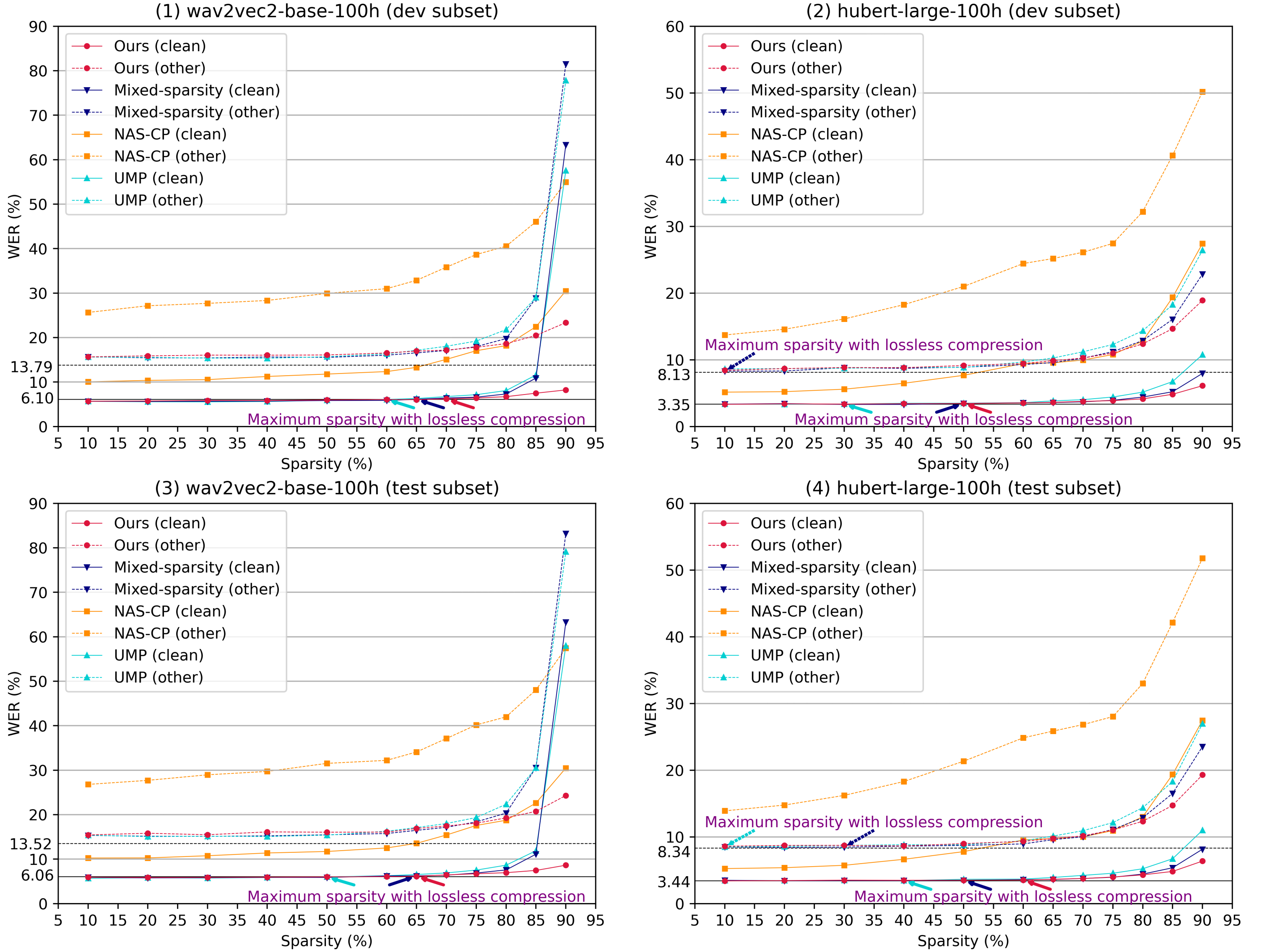}
    \vspace{-2mm}
    \caption{The ASR performance of the pruned wav2vec2-base-100h on the (1) dev and (3) test subsets, as well as the pruned hubert-large on the (2) dev and (4) test subsets with different sparsity using different methods. Abbreviations are the same as those in Figure~\ref{frame}. Color-matched arrows point to the maximum sparsity preserving lossless compression of the methods in corresponding color.}
    \vspace{-6mm}
    \label{w2v-dev}

\end{figure*}








\vspace{-0.3cm}
\section{Experiments}
\vspace{-0.1cm}
\subsection{Experimental setup}
\vspace{-0.1cm}
\noindent\textbf{Uncompressed baselines and data.}
For wav2vec2.0-\textit{base}, the wav2vec2-base-100h is downloaded from Huggingface\footnote{\href{https://huggingface.co/facebook/wav2vec2-base-100h}{Huggingface: facebook/wav2vec2-base-100h}\label{hf}} as our baseline. For HuBERT-\textit{large}, we fine-tuned HuBERT-large-ll60k\footnote{\href{https://huggingface.co/facebook/HuBERT-large-ll60k}{Huggingface: facebook/HuBERT-large-ll60k}} for 20 epochs as our baseline, with other setups consistent with those in \textbf{One-pass pruning and fine-tuning}. All systems are trained on LibriSpeech's~\cite{panayotov2015librispeech} 100-hour clean set. 

\noindent\textbf{One-pass pruning and fine-tuning.}
We utilized the AdamW optimizer with a learning rate of 2e-4 \timnew{and a batch size of 32} for both wav2vec2.0 and HuBERT systems. 
A linear warmup is implemented for the first 10\% of the training steps, followed by a linear decay to zero. All pruned wav2vec2.0-\textit{base} and HuBERT-\textit{large} systems are obtained by fine-tuning 30 and 10 epochs from the baselines of wav2vec2.0-\textit{base} and HuBERT-\textit{large}, respectively.
The threshold $t$ in each layer is initialized to 1e-5. $T$ and $\tau$ are cosine-annealed from 0.5 to 0.01.
All experiments are conducted on a single NVIDIA A40 (48 GB).

\vspace{-0.3cm}
\subsection{Main results}

\begin{table}[h]
      \caption{
      WER(\%) of pruned wav2vec2.0-base and HuBERT-large with different sparsity, using fine-tuning epochs of 30 and 10, respectively. Only selected key results are shown here.
      $\dag$ means it has no statistically significant (MAPSSWE~\cite{gillick1989some}, $\alpha$=0.05) WER increase with the corresponding baseline. }
    \vspace{-0.3cm}
    \centering
    \setlength\tabcolsep{6pt}
    \resizebox{\linewidth}{!}{
    \begin{tabular}{c|c|c|c|cc|cc}
        \hline
        \hline
        \multirow{2}{*}{Sys.} & \multirow{2}{*}{Sparsity}
        & \multirow{2}{*}{\shortstack{\# Params\\(M)}} & \multirow{2}{*}{\shortstack{\# Additional\\components}} & \multicolumn{2}{c|}{dev} & \multicolumn{2}{c}{test}\\
        \cline{5-8}
         & & & &  clean & other & clean & other\\
         \hline
         \multicolumn{8}{l}{\textbf{Baselines}}\\
         \hline
         1 &wav2vec2.0-\textit{base} 
 & 95.04 &\multirow{2}{*}{-} &6.10 	&13.79 &	6.06 &	13.52\\
         2 & HuBERT-\textit{large}  & 316.60 &  &3.35 &	8.13 	&3.44 	&8.34 \\
         \hline
         
         \multicolumn{8}{l}{\textbf{Pruned wav2vec2.0-\textit{base} systems using Sparsity-aware Self-pinching Gates}}\\
          \hline
         3 & 50\%  & 51.93 &\multirow{8}{*}{72}  &5.97$^{\dag}$ &	16.09	&6.00$^{\dag}$ 	&16.06 \\
         4 & 60\% & 43.44 &  &6.06$^{\dag}$ &	16.52 &	6.07$^{\dag}$ 	&16.08 \\
         5 & \hnxu{65\%} & 39.19 & &6.03$^{\dag}$& 	16.97 &	\textbf{6.12$^{\dag}$} &	16.89 \\
         6 & 70\%  & 34.94 & &\textbf{6.21$^{\dag}$} &	17.17 	&6.45 &	17.46  \\
         7 & 75\% & 30.70 & &6.41 	&17.82 &	6.71 &	18.08 \\
         8 & 80\% & 26.45 &  &6.71 	&18.63 	&6.95 	&19.27 \\
         9 & 85\% & 22.20 &  & 7.49 	&20.47 &	7.46 	&20.74 \\
         10 & 90\% & 17.96 & &8.22 &	23.36 &	8.63 	&24.30 \\
         \hline
         \multicolumn{8}{l}{\textbf{Pruned HuBERT-\textit{large} systems using Sparsity-aware Self-pinching Gates}}\\
         \hline
         11 & 50\%  &164.37 &\multirow{8}{*}{144} &\textbf{3.45$^\dag$} 	&9.14 	&3.51$^\dag$ 	&8.99  \\
         12 & 60\% &134.14  & &3.54 	&9.44 	&\textbf{3.54$^\dag$} 	&9.33  \\
         13 & \hnxu{65\%}  &119.03  & &3.62 &	9.86 &	3.65 &	9.75 \\
         14 & 70\% &103.92  &  &3.76 &	10.27 &	3.76 	&10.15 \\
         15 & 75\% &88.81  &  &3.88 	&11.01 &	3.98 &	11.04 \\
         16 & 80\%  &73.70 & &4.19 	&12.51 	&4.36 	&12.39 \\
         17 & 85\%  &58.59 & &4.83 &	14.66 &	4.82 &	14.72 \\
         18 & 90\% &43.48 &  &6.11 	&18.91 &	6.37& 	19.29 \\
         \hline
         \multicolumn{8}{l}{\textbf{Pruned HuBERT-\textit{large} systems using Mixed-sparsity}}\\
         \hline
         19 & 10\% &285.25 &\multirow{4}{*}{-} &3.35$^\dag$ 	&\textbf{8.29$^\dag$} 	&3.49$^\dag$ 	&8.48$^\dag$ \\
         20 & 20\%  & 255.03 & &3.43$^\dag$ &	8.32 	&3.43$^\dag$ 	&8.42$^\dag$ \\
         21 & 30\% & 224.81 & &3.31$^\dag$ 	&8.81 	&3.45$^\dag$ 	&\textbf{8.47$^\dag$} \\
         \cline{1-3} \cline{5-8}         22 & 50\% & 164.37 &  & 3.46$^\dag$ &	8.89 &3.49$^\dag$ 	&8.72 \\
         \hline
         \multicolumn{8}{l}{\textbf{Pruned HuBERT-\textit{large} systems using NAS-based Channel-wise Pruning}}\\
         \hline
         23 & 50\% & 164.37 & 1008  &7.67 	&20.99 &	7.79 &	21.34  \\
         \hline
         \multicolumn{8}{l}{\textbf{Pruned HuBERT-\textit{large} systems using Uniform Magnitude-based Pruning}}\\
         \hline
         24 & 50\% & 164.37 & -  &3.49 	&8.85 	&3.64 	&8.83 \\
         \hline         
    \end{tabular}
        }
    \vspace{-0.7cm}
    \label{Ours-wav2vec}

\end{table}

To facilitate comparison, we implemented Uniform Magnitude-based Pruning (UMP), NAS-based Channel-wise Pruning (NAS-CP)\footnote{Empirically, in NAS-CP, $\eta$ is set to 4e-5 and 3e-5 for wav2vec2.0-\textit{base} and HuBERT-\textit{large}
\timnew{, respectively, when the desired sparsity is less than 75\%; otherwise, it is set to 2e-4 and 5e-5}.} and our method (Ours)\footnote{Empirically, in Ours, $\eta$ is set to 2e-5 and 1e-6 for wav2vec2.0-\textit{base} and HuBERT-\textit{large}, respectively, when the desired sparsity is less than 65\%; otherwise, it is set to 3e-5 and 2e-6.},
as shown in Fig.~\ref{w2v-dev}.
\vspace{-0.3cm}
\subsubsection{Comparison with Uniform Magnitude-based Pruning}

Pruned wav2vec2.0-\textit{base} using both UMP and our method achieve lossless compression with all sparsity of below 50\% on the clean subsets. 
Beyond 50\%, pruned wav2vec2.0-\textit{base} and HuBERT-\textit{large} using ours consistently exceed those using UMP (Fig.~\ref{w2v-dev} (1)-(4)) on all subsets, respectively. On the dev-clean subset, pruned wav2vec2.0-\textit{base} and HuBERT-\textit{large} using ours achieve lossless compression with maximum sparsity of 70\% (Fig.~\ref{w2v-dev} (1) and Sys.~6 in Tab.\ref{Ours-wav2vec}) and 50\% (Fig.~\ref{w2v-dev} (2) and Sys.~11 in Tab.\ref{Ours-wav2vec}), respectively, versus UMP's 60\% and 30\% (Fig.~\ref{w2v-dev} (1)-(2)). On the test-clean subset, ours achieves 65\% (Fig.~\ref{w2v-dev} (3) and Sys.~5 in Tab.\ref{Ours-wav2vec}) and 60\% (Fig.~\ref{w2v-dev} (4) and Sys.~12 in Tab.\ref{Ours-wav2vec}), compared to UMP's 50\% and 40\% (Fig.~\ref{w2v-dev} (3)-(4)).


\zq{We conjecture that the inferior performance of UMP is partially caused by the inconsistency that decouples pruning from parameter optimization. To verify this, we implement a decoupled version of our method, which involves pruning the parameters with relatively small magnitude from the uncompressed model to meet the layer-wise sparsity obtained from our one-pass method, as shown in Fig.~\ref{w2v-dev} under the label \textit{Mixed-sparsity}.}
It outperforms UMP across most sparsity levels, suggesting that our one-pass method effectively accounts for the varying sensitivities of different layers.
However, it performs worse than our one-pass method, highlighting that the inconsistency between pruning and parameter update negatively affects ASR performance.
Notably, on the less-explored dev-other and test-other subsets, pruned HuBERT-\textit{large} systems using \textit{Mixed-sparsity} maintain lossless compression with maximum sparsity of 10\% (Fig.~\ref{w2v-dev} (2) and Sys.~19 in Tab.\ref{Ours-wav2vec}) and 30\% (Fig.~\ref{w2v-dev} (4) and Sys.~21 in Tab.\ref{Ours-wav2vec}), respectively, while UMP only achieves 10\% on the test-other subset (Fig.~\ref{w2v-dev} (4)). 

\vspace{-0.3cm}
\begin{table}[h]
\caption{The ASR performance comparison of our method versus previous compression methods on the test-clean subset. The values in brackets ``('' and ``)'' represent the transformer sparsity. Comp.ratio: the compression ratio relative to the uncompressed model. *: from leaderboard of SUPERB~\cite{superb}.
} 
    \vspace{-0.3cm}
    \centering
    \setlength\tabcolsep{1pt}
    \resizebox{\linewidth}{!}{
    \begin{tabular}{c|c|c|c|c|c|c|c|c}
        \hline
        \hline
        \multirow{2}{*}{Sys.} & \multirow{2}{*}{System} & \multirow{2}{*}{\shortstack{Train\\set}} & \multirow{2}{*}{\shortstack{\# Params\\(M)}} & \multirow{2}{*}{\shortstack{Comp.\\ratio}} & \multicolumn{3}{c|}{Training time \& config} & \multirow{2}{*}{WER\timnew{\%}}\\
        \cline{6-8}
         & & & & & Hours& Epochs& Iterations&\\
         \hline
         \multicolumn{9}{l}{\textbf{Baselines}}\\
         \hline
        1 & wav2vec2.0-\textit{base} &\multirow{4}{*}{100h} & 95.04 &\multirow{4}{*}{-} &\multirow{4}{*}{-} &\multirow{4}{*}{-}&\multirow{4}{*}{-} &6.06 \\
        2 & HuBERT-\textit{base} & &94.68 & & & & &6.42* \\
        3 & WavLM-\textit{base+} & &94.70 & & & & &\hnxured{5.59*} \\
        4 & HuBERT-\textit{large} & &316.60 & & & & &3.44 \\
         \hline
         \multicolumn{9}{l}{\textbf{Prior Compression Methods}}\\
         \hline
         5 & DistilHuBERT~\cite{distilhubert} &960h &23.49&4.03 &55 &200 &- &13.37* \\
         \hline
         6 & FitHuBERT-100~\cite{fithubert} &\multirow{2}{*}{100h} &\multirow{2}{*}{22.49} &4.21 &$>$12 &\multirow{2}{*}{100} &-  &12.66* \\
        7 & FitW2V2-100~\cite{fithubert} & & &4.23 &-& &- &14.77* \\
        \hline
         8 & FitHuBERT-960~\cite{fithubert} &\multirow{2}{*}{960h} &22.49 &4.21 &$>$93.6 &\multirow{2}{*}{80} &- &12.09* \\
        9 & FitW2V2-960~\cite{fithubert} & &31.63 &3.00 &-& &- &11.44* \\
        \hline
        10 & 12-Layer Half~\cite{deepvswide} &960h &26.87 &3.52 &- &-& 200K &10.96*\\
        \hline
        11 & DPHuBERT~\cite{peng2023dphubert} & \multirow{2}{*}{960h} &23.59 &4.01 & \multirow{2}{*}{24} &- &\multirow{2}{*}{75K} &10.47*\\
        12 & DPWavLM~\cite{peng2023dphubert} &  &23.59 &4.01 & &- & &10.19*\\
        \hline
        13 & SKHuBERT~\cite{zampierin2024skillsimilarityawareknowledgedistillation} & \multirow{2}{*}{960h} &23.59 &4.01 & \multirow{2}{*}{24} &- &\multirow{2}{*}{75K} &10.78*\\
        14 & SKWavLM~\cite{zampierin2024skillsimilarityawareknowledgedistillation} &  &23.51 &4.03 & &- & &10.03*\\
        \hline
        15 & Wang et al.~\cite{wang2023task} &960h &26.57 &3.56 &36 &- & 200K &10.29*\\
        \hline
        16 & DKD LSTM HuBERT~\cite{de2023lstmhubert} &960h &18.80 &5.04 &- &- & 200K &10.64*\\
        \hline
        17 & SparseWAV2VEC2 (85\%)~\cite{gu2024sparsewav} &\multirow{3}{*}{100h} &22.27  &4.27 &- &\multirow{3}{*}{$>$40} &- & 8.08\\
        18 & SparseWAVLM+ (85\%)~\cite{gu2024sparsewav} & &22.98  &4.12 &- & &- & 7.12\\
        19 & SparseWAV2VEC2 (75\%)~\cite{gu2024sparsewav} & &-  &- &- & &-  &7.11 \\
        \hline
         \multicolumn{9}{l}{\textbf{Ours}}\\
         \hline
         20 & \hnxu{Pruned wav2vec2-\textit{base} (85\%)} &\multirow{2}{*}{100h}  &\textbf{22.20} &\textbf{4.28} & \multirow{2}{*}{\textbf{11}} &\multirow{2}{*}{\textbf{30}} &\multirow{2}{*}{\textbf{27K}}&\textbf{7.46}\\
         21 & Pruned wav2vec2-\textit{base} (75\%) & &30.70 &3.10 & & & &6.71 \\
         \hline
         22 & \hnxu{Pruned WavLM-\textit{base+} (85\%)} & 100h& 22.21&4.26 &\textbf{11} &\textbf{30} &\textbf{27K} &\textbf{7.05}\\
         \hline
         23 & Pruned HuBERT-\textit{large} (90\%)&\multirow{1}{*}{100h} &43.48 &7.28 & \multirow{1}{*}{\textbf{13}} &\multirow{1}{*}{\textbf{10}} &\multirow{1}{*}{\textbf{9K}} & \textbf{6.37}\\
         \hline
         \hline

    \end{tabular}      
    }
    \vspace{-0.4cm}
    \label{comparison}

\end{table}

\vspace{-0.3cm}
\subsubsection{Comparison with NAS-based Channel-wise Pruning}
\vspace{-0.1cm}


Predictably, compared with NAS-CP that prunes multiple channels, our method applied to both wav2vec2.0-\textit{base} and HuBERT-\textit{large} achieves consistently lower WERs across all sparsity as demonstrated in Fig.~\ref{w2v-dev} (1)-(4).
Regarding the additional components, 
NAS-CP requires 7 architecture-dependent parameters per layer (e.g., 1008 total for HuBERT-\textit{large} of Sys.~23 in Tab.~\ref{Ours-wav2vec}) in our configurations (Sec.~\ref{NAS}). In contrast, our method maintains only one threshold per layer (e.g., 144 total for HuBERT-\textit{large} of Sys.~11 in Tab.~\ref{Ours-wav2vec}).


\vspace{-0.3cm}
\subsection{Comparison with other compression methods}
\vspace{-0.2cm}

As illustrated in Tab.~\ref{comparison}, our method (Sys.~20) achieves $\geq$2\% absolute WER reduction over Sys.~5-16 under equivalent model-size limits. In order to compare training times, we consolidated different types (Hours/Epochs/Iterations) of training-time metrics from previous works into standardized comparisons, where our method demonstrates consistent time-efficiency advantages. Notably, the WERs of Sys.~20 and Sys.~21 using our method surpass Sys.~17 and Sys.~19, respectively, while requiring at least 25\% less fine-tuning epochs, highlighting the effectiveness and efficiency of our approach. 
\hnxured{We also pruned WavLM-\textit{base+} downloaded from Huggingface\footnote{Huggingface: patrickvonplaten/wavlm-libri-clean-100h-base-plus} using the same setup as wav2vec2.0-\textit{base}, yielding the lowest WER (Sys.~22 in Tab.~\ref{comparison}) of 7.05\% on the test-clean subset under the model-size constraint of 23M parameters, compared to Sys.~5-19 in Tab.~\ref{comparison}.} 

\vspace{-0.3cm}
\section{Conclusion}
\vspace{-0.1cm}
We introduced a cutting-edge one-pass compression method that simultaneously prunes and trains SSL speech foundation models using a threshold per layer. 
Our method shows superior performance under the same model-size constraint while reducing the fine-tuning time compared to the previous works. \hnmod{We will further apply our methods to more datasets in future work.}

\section{Acknowledgements}
This research is supported by Hong Kong RGC GRF grant No. 14200220, 14200021, 14200324 and Innovation Technology Fund grant No. ITS/218/21.

\bibliographystyle{IEEEtran}
\bibliography{mybib}

\end{document}